# Adapting Behaviour Based On Trust In Human-Agent Ad Hoc Teamwork[*]

Ana Carrasco[**]

INESC-ID, Instituto Superior Técnico, University of Lisbon
Lisbon, Portugal
ana.carrasco@gaips.inesc-id.pt

**Abstract.** This work proposes a framework that incorporates trust in an ad hoc teamwork scenario with human-agent teams, where an agent must collaborate with a human to perform a task. During the task, the agent must infer, through interactions and observations, how much the human trusts it and adapt its behaviour to maximize the team's performance. To achieve this, we propose collecting data from human participants in experiments to define different settings (based on trust levels) and learning optimal policies for each of them. Then, we create a module to infer the current setting (depending on the amount of trust). Finally, we validate this framework in a real-world scenario and analyse how this adaptable behaviour affects trust.

## 1 Introduction

Ad Hoc Teamwork (AHT) [8] is a well-known research problem where an agent must be able to cooperate with a group of teammates (humans or agents), without any prior coordination or communication protocols. The AHT is characterized by three assumptions: the agent has no prior coordination or communication mechanisms established with its teammates, the agent has no control over the teammates and all teammates are assumed to have a common objective (or task) [5]. As defined by Melo and Sardinha [4], the AHT problem can be divided into three sub-tasks: identifying the common task, identifying the teammates and what they are doing and plan actions accordingly.

A review of the recent literature on the subject reveals that, as autonomous agents capabilities advance, human-robot collaboration scenarios become more common. We argue that AHT scenarios with human-agent teams are especially relevant, since they solve the problem of integrating an ad hoc agent in a team with a human, without requiring any prior experience with that human. However, very few works on AHT focus on teams with humans, and even fewer regard the human-agent trust dynamics. Ribeiro et al. [6] introduce a framework for AHT in human-robot teams, tackling two main challenges in human-robot collaboration: task-related challenges and communication challenges between robots and humans. This work builds on that by identifying and addressing the challenges raised by the trust dynamics between the human and the agent.

[*] This work was partially supported by national funds through FCT, Fundação para a Ciência e a Tecnologia, under project UIDB/50021/2020 (INESC-ID multi-annual funding) and the HOTSPOT project, with reference PTDC/CCI-COM/7203/2020. In addition, this material is based upon work supported by TAILOR, a project funded by EU Horizon 2020 research and innovation programme under GA N. 952215.

[**] The advisors of this work are: Ana Paiva (paiva.a@gmail.com), Alberto Sardinha (jose.alberto.sardinha@tecnico.ulisboa.pt), Francisco S. Melo(fmelo@inesc-id.pt)

Trust in human-robot interaction is a well researched topic. Lee and See [3] and Schaefer [7] explore the impact of trust in human-robot interactions, how to model and measure it. These works argue that successful collaboration between humans and robots require appropriate levels of mutual trust, since that trust impacts the task performance. Chen et al. [1] introduce a computational model based on a POMDP that uses trust as a latent variable. This model allows a robot to infer the trust of a human and choose actions that maximize the long term performance. This model is validated through human participant experiments on a table-clearing task.

Inspired by these works, the goal of this work is to introduce a framework, in a real world AHT scenario, where the agent is able to infer how much the human trusts it and change the way it cooperates with the human. We hypothesize that the agent's adaptable behaviour will not only improve task performance but also calibrate the amount of trust the human has on the agent.

## 2   Experimental Setup

The structure for the proposed framework is depicted in Figure 1.

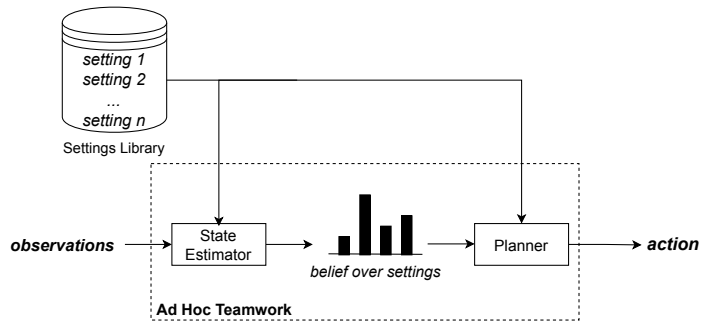

**Fig. 1.** Structure of the proposed framework.

First, information on different settings, i.e., different models of human behaviour based on the level of trust they show towards the agent, is saved into a *settings library*. The agent then infers the current setting by combining this information with observations from the environment and the teammate. It then updates a distribution of possible settings (*beliefs*), and uses these *beliefs* and the information about the setting to choose the best action to take.

To validate this framework, we propose a controlled experiment where human participants must complete a task with an agent. The structure of this experiment is inspired by the work of Ribeiro et al. [6] and Chen et al. [1]. In the work of Ribeiro et al. [6], the robot assists the human in a task that follows a set of rules under the Toxic Waste Domain. In this domain, the robot and the human must clean the toxic waste from a room. The human must pick up the waste and dispose of it while the robot's role is to be the container where the waste is disposed of. Due to the toxicity of the waste, the human is penalized for the time holding the waste.

This work takes this scenario but adjusts the rules and the environment characteristics to add some risk related to trusting the agent. This allows for an assessment of the human's trust in the agent (i.e., if and when that trust is higher or lower), similarly to what was done by Chen et al. [1]. In their work, the risk was associated with the robot dropping items, whereas in this experiment the risk is directly connected to time consumption. The experiment is timed, and the participant is penalized for taking longer to complete the task. Note that, despite the human's perception, the robot is capable of executing its part of the task perfectly by itself. The performance of the team depends on how much the human trusts in the robot to execute the task in a reasonable time period. If implemented correctly, the robot's adaptable behaviour will allow for an improvement on this performance, keeping up with the natural evolution of the human-robot trust dynamics.

The first stage of the study consists of gathering data regarding the behaviour of the participants when collaborating with the agent. This will allow us to choose the best way to measure trust levels and build the *settings library* to be used in the framework. in this first stage, the proposed experimental setup is tested virtually in an environment called *Overcooked*[1]. In this scenario, we introduce elements like an *"icy zone"*, where the agent's navigation capabilities are decreased. The goal is to simulate some level of risk and create some level of uncertainty regarding the robot's performance, which is an inherent trait in the real-world scenario, but might lack in the virtual one.

## 3  Future Work

Currently, the study is still in its early stages of modeling the experiment environment and gathering data. The next steps include exploring the best way to measure trust in the proposed scenario (whether subjectively, objectively or both [2]), in order to develop a trust inferring module; Compute optimal policies adequate for different levels of trust; Test the agent with adaptable behaviour in a real world scenario; Study the impact of the robot's adaptable behaviour in trust, during and after the task.

---

[1] https://github.com/HumanCompatibleAI/overcooked_ai